\documentclass[12pt]{iopart}

\usepackage{graphicx}
\usepackage{epstopdf}
\usepackage{bm}
\DeclareGraphicsExtensions{.JPG,.eps}

\begin{document}

\title[Fast plasma dilution in ITER with pure Deuterium Shattered Pellet Injection]{Fast plasma dilution in ITER with pure Deuterium Shattered Pellet Injection}

\author{E. Nardon$^1$, D. Hu$^2$, M. Hoelzl$^3$, D. Bonfiglio$^4$ and the JOREK team$^a$}

\address{$^1$CEA, IRFM, F-13108 Saint-Paul-lez-Durance, France}
\address{$^2$School of Physics, Beihang University, Beijing 100191, China}
\address{$^3$Max Planck Institute for Plasma Physics, Boltzmannstr. 2, 85748 Garching b. M., Germany}
\address{$^4$Consorzio RFX, Corso Stati Uniti 4, 35127 Padova, Italy}
\address{$^a$URL: https://www.jorek.eu for a current list of team members}

\ead{eric.nardon@cea.fr}
\vspace{10pt}

\begin{abstract}

JOREK 3D non-linear MagnetoHydroDynamic (MHD) simulations of pure Deuterium Shattered Pellet Injection in ITER are presented. Considering a 15 MA L-mode plasma with a thermal energy content of 36 MJ from the non-activation phase of ITER operation, it is shown that such a scheme could allow diluting the plasma by more than a factor 10 without immediately triggering large MHD activity, provided the background impurity density is low enough. This appears as a promising strategy to reduce the risk of hot tail Runaway Electron (RE) generation and to avoid RE beams altogether in ITER, motivating further studies in this direction. 

\end{abstract}

%
%
%
%
%

\section{Introduction}

Shattered Pellet Injection (SPI) is the reference concept for the ITER Disruption Mitigation System (DMS) \cite{Lehnen_IAEA_2018}\cite{Lehnen_JNM_2015}. One of the main objectives of the DMS is the avoidance of large Runaway Electron (RE) \cite{Breizman_2019} beams. There are two particular concerns regarding RE beam formation during ITER disruptions \cite{Boozer_PoP_2015}\cite{Hender_NF_2007}. The first one is that the hot tail mechanism \cite{Smith_PoP_2005} may be very efficient when a hot plasma disrupts. This is suggested by recent experiments on DIII-D \cite{Paz-Soldan_2020} but remains largely an open question due to our limited understanding of the Thermal Quench (TQ) phase of a disruption. The second concern is that, even if the generation of RE seeds (whether by the hot tail mechanism or by other mechanisms such as Dreicer, Tritium $\beta$ decay or Compton scattering of gamma rays emitted by the activated wall) is small, these seeds may still be converted into a large RE beam via the avalanche mechanism \cite{Rosenbluth_NF_1997} during the current quench phase of the disruption \cite{Boozer_PPCF_2019}. Suppressing the avalanche would require raising the electron density $n_e$ up to the so-called Rosenbluth density, which is in the range of several times $10^{22}$ m$^{-3}$ \cite{Hender_NF_2007}. This would require an increase in $n_e$ by more than 2 orders of magnitude, which seems to be a difficult task. Even if the Rosenbluth density cannot be reached, increasing $n_e$ may still be useful by raising the critical energy for running away and thus reducing seed REs, with the exception of Compton seeds \cite{Martin-Solis_NF_2017}\cite{Boozer_PPCF_2019}. Modelling studies suggest that, assuming a moderate hot tail seed, large RE beams may be avoided during ITER disruptions if $n_e$ could be raised by a factor of 20 to 40 throughout the plasma \cite{Martin-Solis_NF_2017}\cite{Aleynikov_ITER_report}, although more recent work puts this in question \cite{Vallhagen_2020}. It is imagined that this increase in $n_e$ could be achieved by injecting shattered pellets containing a large quantity of Deuterium (D$_2$) mixed with a much smaller quantity of impurities like Argon or Neon, the latter being necessary to mitigate thermal and electromagnetic loads \cite{Lehnen_IAEA_2018}. However, it is not established that such large amounts of material could actually be assimilated by the plasma, and even less so in a uniform fashion.

An important point is that the ablation of pellet shards is strongly dependent on the electron temperature $T_e$ \cite{Pegourie_PPCF_2007}. With typical ITER SPI parameters, ablation is strong at large (multi-keV) pre-TQ $T_e$, but becomes almost negligible at low ($\sim 10$ eV or less) post-TQ $T_e$. In order to maximize the amount of ablated material, a late TQ thus appears clearly beneficial. 

A problem when using pellets containing impurities is that they typically trigger a TQ while shards are still far away from the plasma center \cite{Shiraki_PoP_2016}. This effect, which is captured by 3D non-linear MHD modelling \cite{Kim_PoP_2019}, is due to the fact that, via radiation, impurities cool the electrons down to $T_e \sim 10$ eV or below on a timescale which is much shorter than the time it takes for shards to reach the plasma center, $\tau_{travel}  \equiv v_{s}/a$, where $v_s$ is the shards velocity and $a$ the plasma minor radius. This $T_e$ drop increases the plasma resistivity $\eta$ (which scales like $T_e^{-3/2}$) dramatically, which in turn modifies the current density distribution \cite{Artola_PoP_2020}, destabilizes MHD modes and leads to the TQ. 

On the other hand, pure D$_2$ SPI may behave differently because Deuterium radiates very little. Electron cooling in this case comes essentially from dilution. If the dilution factor is of the order of, say, 30, and if the initial $T_e$ is, say, 3 keV, then the post-dilution $T_e$ is 100 eV. The effect on $\eta$ is thus much more limited than in the presence of impurities, such that the current density distribution will evolve on a much slower timescale. As a matter of fact, it has been observed on DIII-D \cite{Shiraki_PoP_2016} and JET \cite{Lehnen_private_2020} that the cooling time, defined as the duration between the moment when first shards reach the edge of the plasma and the onset of the TQ, is much longer when pure D$_2$ is used than when the pellet contains impurities. Also, studies on rapid plasma shutdown by massive uniform D$_2$ delivery in DIII-D and ITER with NIMROD found that the plasma remains stable to $n=1$ and $n=2$ modes ($n$ being the toroidal mode number)  \cite{Izzo_2009}.


These considerations suggest that pure D$_2$ SPI may be much more efficient at raising $n_e$ than mixed D$_2$-impurities SPI, because the plasma is likely to remain hot for a longer duration, allowing a much more efficient ablation. Also, the longer cooling time will allow shards to penetrate deeper and provide direct core fuelling. As already mentioned above, impurities are necessary to mitigate thermal and electromagnetic loads, but these could be injected in a second step, after the pure D$_2$ SPI, by means of a second SPI. Results shown in this paper suggest that there would be sufficient time for this.

Preliminary axisymmetric simulations by Akinobu Matsuyama with the 1.5D transport code INDEX \cite{Matsuyama_REM_2020} support the above considerations by showing that with pure D$_2$ SPI, the current density profile is weakly affected, at least on a $\tau_{travel}$ timescale. In the present work, we use the JOREK 3D non-linear MHD code \cite{Huijsmans_NF_2007}\cite{Czarny_JCP_2008}\cite{Hoelzl_JOREK_2020} to further assess the feasibility of this scheme. 

The paper is constructed as follows. In Section \ref{sec:setup}, we introduce the simulation setup. Section \ref{sec:axisym_sims} presents axisymmetric simulations and discusses the impact of the resistivity used in the simulation and of background impurities. Section \ref{sec:3D_sims} describes 3D simulations and addresses in particular the effect of the toroidal localization of the ablation clouds and (again) of background impurities. Finally, Section \ref{sec:discussion} discusses the results and concludes.

\section{Simulation setup}
\label{sec:setup}

The simulations presented in this study were performed with the fixed-boundary reduced MHD version of JOREK comprising an extension for neutrals and SPI. The model is described in detail in \cite{Hu_NF_2018}. The same tool has been used recently to simulate pure D$_2$ SPI in JET \cite{Hu_NF_2018} and ASDEX Upgrade \cite{Hoelzl_SPI_2020}.

\begin{figure}[ht]
	\centering
	\includegraphics[width=50mm]{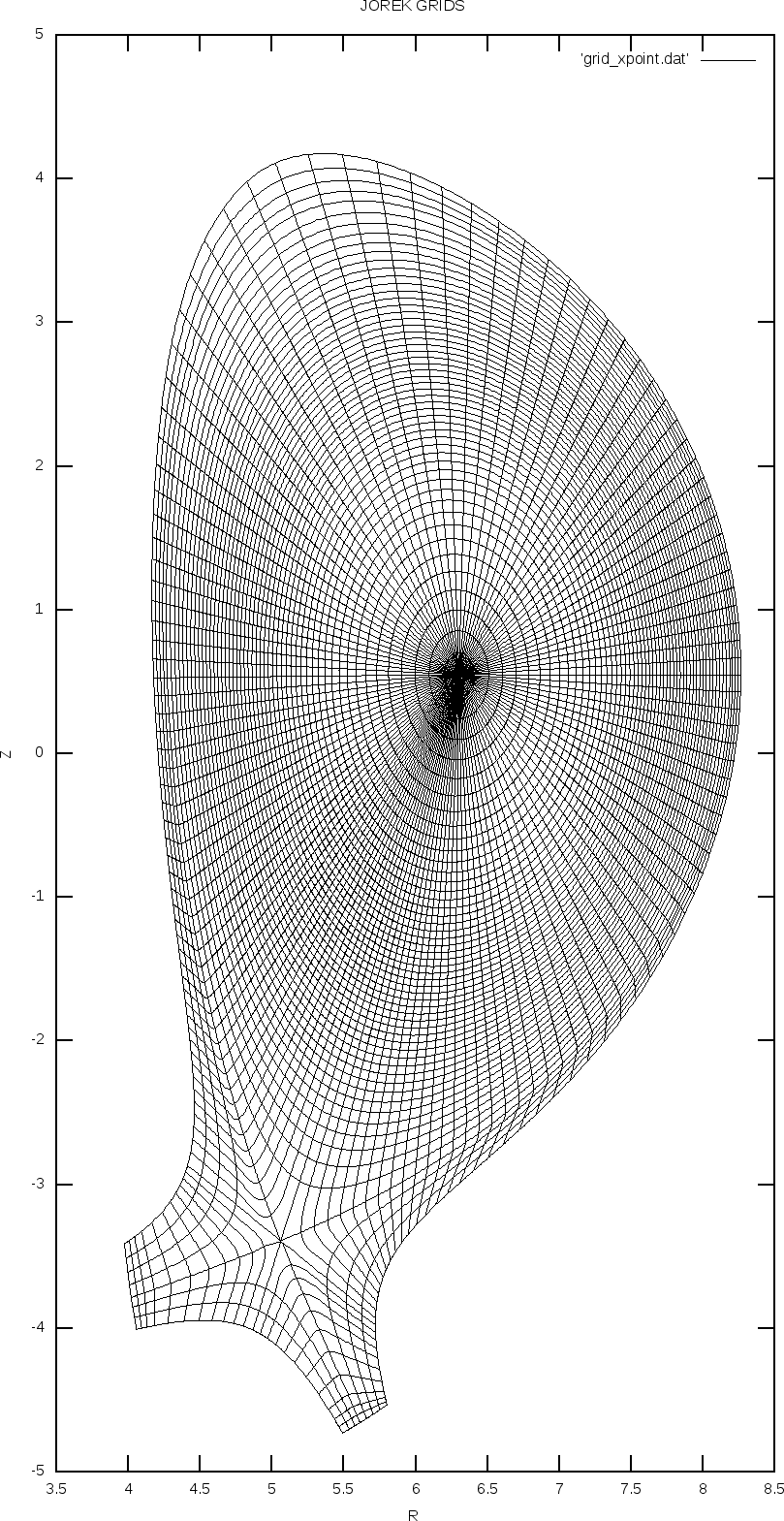}
	\caption{Computational grid used in the JOREK simulations}
	\label{Fig_grid}
\end{figure}

The target plasma used for this study is taken at time $t=60$ s from a CORSICA simulation of an ITER Hydrogen L-mode scenario (Scenario 5 from Table 1 in \cite{SH_Kim_ref_scen}). The plasma current is 15 MA, the toroidal magnetic field 5.3 T and the thermal energy content 36 MJ. The computational grid in the poloidal plane is shown in Fig. \ref{Fig_grid}. It comprises about 6000 elements. In the toroidal direction, JOREK uses a Fourier decomposition. Axisymmetric simulations shown in Section \ref{sec:axisym_sims} include only the $n=0$ component whereas the 3D simulations of Section \ref{sec:3D_sims} include harmonics from $n=0$ up to $n=10$.

\begin{figure}[ht]
	\centering
	\includegraphics[width=100mm]{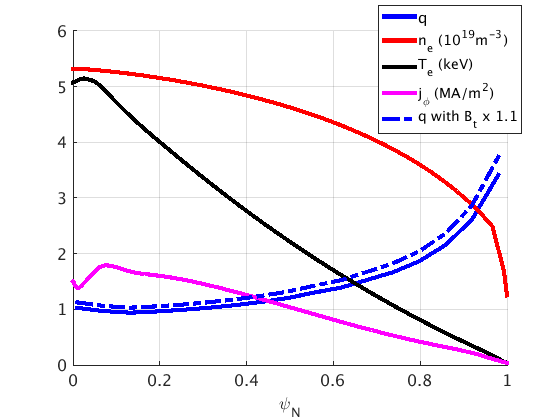}
	\caption{Initial profiles for the target ITER plasma used in this study (15 MA Hydrogen L-mode)}
	\label{Fig_init_profs}
\end{figure}

Hydrogen is replaced by Deuterium in JOREK since the model assumes that neutrals and ions are of the same species. This is not expected to have important consequences since the ion mass should not play an important role here. Initial profiles of the safety factor $q$, electron density $n_e$, electron temperature $T_e$ and toroidal current density $j_\phi$, as a function of the normalized poloidal magnetic flux $\psi_N$ ($\psi_N=0$ on the magnetic axis and $\psi_N=1$ at the last closed flux surface), are shown in Fig. \ref{Fig_init_profs}. Central values of $n_e$ and $T_e$ are $5.3 \times 10^{19}$ m$^{-3}$ and 5 keV, respectively. It can be noted that $q$ is slightly smaller than 1 near the plasma center. Due to this, the plasma has an intrinsically unstable $n=1$ mode localized in the core, which complicates the interpretation of our simulations. We have therefore increased $B_t$ by 10 $\%$ so as to make $q > 1$ and have an intrinsically stable target plasma. The corresponding $q$ profile is shown as the dash-dotted blue line in Fig. \ref{Fig_init_profs}. 

The injected pellet is a `large' pellet from the present ITER DMS design, which is cylindrical with a diameter $D=28$ mm and a length over diameter ratio $L/D = 2$ \cite{Lehnen_IAEA_2018}. The total number of Deuterium atoms in the pellet is $2 \times 10^{24}$. We assume, for simplicity, the pellet to be shattered into 1000 shards of equal radius (about 2 mm). All shards originate from the position $R=8.25$ m, $Z=0.5$ m, $\phi=0$ (i.e. from the low field side midplane). The velocity vector of each shard has a modulus picked randomly between $100$ and $300$ m/s, and a direction picked randomly inside a cone of aperture 0.349 radians.

The ablation rate $N'$ (i.e. the number of ablated atoms per second) of each shard is calculated according to the neutral gas shielding model (see \cite{Gal_NF_2008} and references therein):

\begin{equation}
	N' = 4.12 \times 10^{16} \cdot r_s^{1.33} \cdot n_e^{0.33} \cdot T_e^{1.64}
\end{equation}

where $r_s$ is the shard radius in m, $n_e$ is in m$^{-3}$ and $T_e$ in eV. The radius of each shard is evolved in time according to its ablation.

Ablated atoms are deposited in an ablation `cloud' via a volumetric source term which has the following spatial shape:
\begin{equation}
S_n\propto\exp\left(-\frac{(R-R_s)^2+(Z-Z_s)^2}{ r_{cloud}^2}\right)\, \exp\left(-\frac{(\phi-\phi_s)^2}{\Delta \phi_{cloud}^2}\right),
\end{equation}
where $(R_s,Z_s,\phi_s)$ corresponds to the shard's position. The poloidal extent of the ablation cloud in our simulations is $r_{cloud} = 20$ cm and the toroidal extent varies between 1 and 1000 radians (the latter corresponding to a virtually axisymmetric cloud). These dimensions are much larger than those of real ablation clouds \cite{Pegourie_PPCF_2007}, but we are constrained by a limited spatial resolution. 

For most simulations, we use a realistic resistivity: $\eta = 2 \times \eta_{Sp} = 5.6 \times 10^{-8} / T_{e}^{3/2}$, where $\eta_{Sp}$ is the Spitzer resistivity \cite{Spitzer_Harm}, $\eta$ and $\eta_{Sp}$ are expressed in $\Omega$.m and $T_e$ in keV. The factor 2 accounts for the fact that only passing electrons carry current (assuming a trapped fraction of 0.5, which is representative of the plasma edge). This corresponds to a Lundquist number $S \equiv \mu_0 a V_A / \eta \simeq 6 \times 10^9$ in the center of the plasma ($a=2$ m is the minor radius and $V_A$ is the Alfv\'en speed). Note that in some simulations, we include background impurity radiation but for simplicity we do not take impurities into account in the resistivity. We use a realistic Spitzer-Harm parallel heat diffusivity \cite{Spitzer_Harm}. The perpendicular heat diffusivity is 10 m$^2$/s. The plasma particle perpendicular diffusivity as well as the isotropic neutral diffusivity are 40 m$^2$/s. Parallel plasma particle transport is purely convective. The central value of the kinematic viscosity acting on the perpendicular flow is equivalent to a diffusivity of 2 m$^2$/s and the viscosity is scaled with temperature in the same way as the resistivity to keep the magnetic Prandtl number approximately constant spatially. The viscosity acting on the parallel flow is constant and equivalent to 200 m$^2$/s. In order to avoid numerical issues related to very low temperatures, Deuterium radiation and recombination are turned off and background impurity radiation is turned off below 5 eV, which has a minor impact on simulations. Small hyperdiffusion terms are also included for numerical reasons but do not affect the results significantly.

\section{Axisymmetric simulations}
\label{sec:axisym_sims}

\begin{figure}[ht]
	\centering
	\includegraphics[width=100mm]{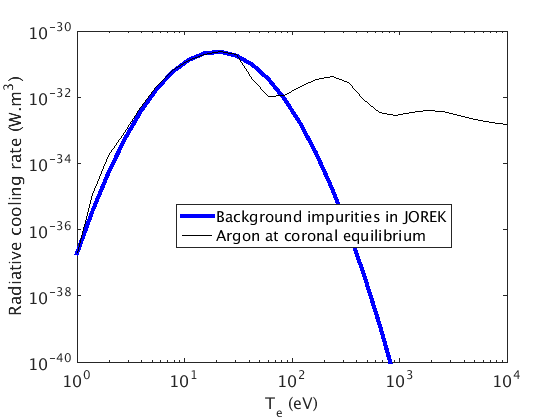}
	\caption{Radiative cooling rate as a function of the electron temperature}
	\label{Fig_Lrad_imp}
\end{figure}

We begin our study with axisymmetric simulations (and thus also with axisymmetric ablation clouds). We present a set of three simulations. The first one has a realistic resistivity, that is $\eta = 2 \times \eta_{Sp}$. This corresponds, in JOREK normalized units, to $\eta_0 = 2 \times 10^{-9}$. The second simulation has a 50 times larger resistivity, that is $\eta_0 = 10^{-7}$ (however, for the Ohmic heating term, the realistic resistivity is kept, so as to avoid spurious Ohmic heating). The third simulation, like the first one, has a realistic resistivity but also includes radiative losses from background impurities with a homogeneous density $n_{imp}=10^{17}$ m$^{-3}$ and a radiative cooling rate resembling that of Argon at coronal equilibrium (at least in the $1-100$ eV range), as shown in Fig. \ref{Fig_Lrad_imp}. 

\begin{figure}[ht]
	\centering
	\includegraphics[width=70mm]{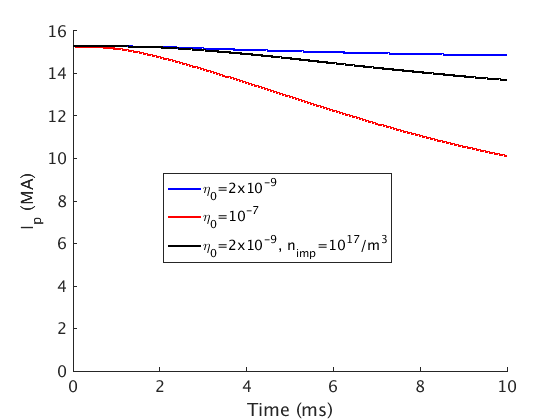}
           \includegraphics[width=70mm]{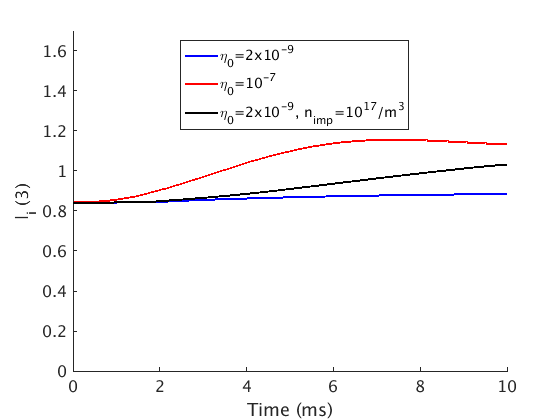}
	\caption{Time evolution of the plasma current $I_p$ (left) and internal inductance $l_i(3)$ (right) for the three axisymmetric simulations: realistic resistivity (blue), resistivity increased by a factor 50 (red), and realistic resistivity with background impurities (black)}
	\label{Fig_axisym_Ip_li3}
\end{figure}

\begin{figure}[ht]
	\centering
	\includegraphics[width=70mm]{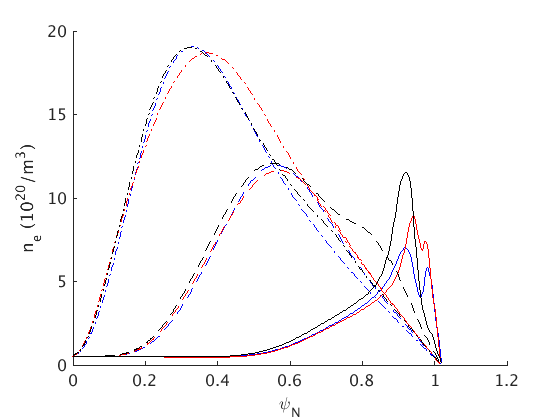}
	\includegraphics[width=70mm]{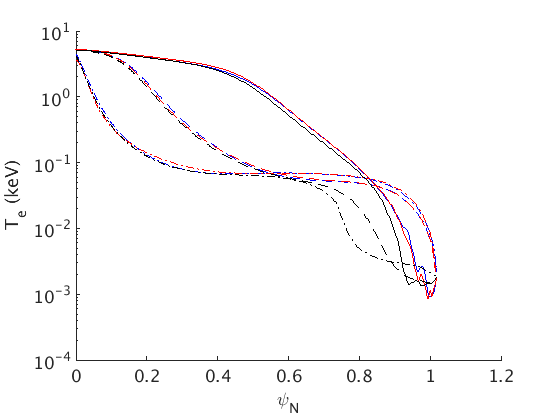}
	\includegraphics[width=70mm]{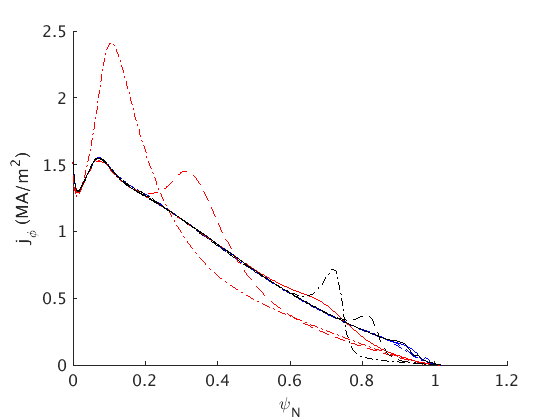}
	\includegraphics[width=70mm]{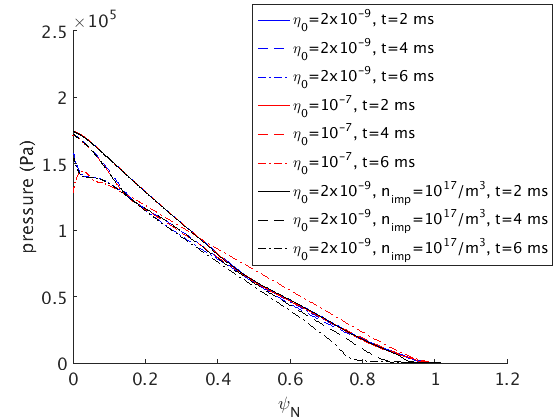}
	\caption{Low field side midplane profiles (as a function of the normalized poloidal flux $\psi_N$) of $n_e$ (top left), $T_e$ (top right), $j_\phi$ (bottom left) and pressure (bottom right) at 2 (plain lines), 4 (dashed lines) and 6 (dash-dotted lines) ms, for the three axisymmetric simulations}
	\label{Fig_axisym_profiles}
\end{figure}

Fig. \ref{Fig_axisym_Ip_li3} shows the time evolution of the plasma current $I_p$ and internal inductance $l_i(3)$, and Fig. \ref{Fig_axisym_profiles} shows profiles of $n_e$, $T_e$, $j_{\phi}$ and pressure at 2, 4 and 6 ms, for the three simulations. Let us consider first the simulation with a realistic resistivity and no background impurities (blue lines). It can be seen in Fig. \ref{Fig_axisym_profiles} that $n_e$ increases massively, first near the edge and gradually towards the core, as shards penetrate deeper and deeper into the plasma (compare the blue $n_e$ profiles at 2, 4 and 6 ms). Due to dilution, $T_e$ decreases accordingly while the pressure profile remains virtually unaffected. In spite of the strong modification of the $n_e$ and $T_e$ profiles, the $j_{\phi}$ profile, and consistently $I_p$ and $l_i(3)$, barely change over a $10$ ms timescale (which corresponds to the typical time needed for shards to reach the plasma center). The second simulation, with the 50 times larger resistivity (red lines), shows virtually the same evolution of the $n_e$, $T_e$ and pressure profiles but, in contrast to the first simulation, displays strong changes in $j_{\phi}$, $I_p$ and $l_i(3)$, demonstrating the key role of the resistivity used in the simulation. It can be seen in the $j_{\phi}$ profile evolution shown in Fig. \ref{Fig_axisym_profiles} that current is removed from the edge and partly re-induced further into the core, with a skin current progressing towards the core, as described in \cite{Artola_PoP_2020}. Accordingly, $I_p$ decreases and $l_i(3)$ increases. The third simulation, with the realistic resistivity and background impurities (black lines), shows a similar $n_e$ and $T_e$ evolution to the other two simulations, with the important difference that $T_e$ is reduced to much lower values in the edge region. This is because, in addition to dilution, electrons are now also cooled by background impurity radiation. The evolution of $j_{\phi}$, $I_p$ and $l_i(3)$ qualitatively resembles that of the second simulation, with current being removed from the edge and re-induced further into the core, but the skin current is sharper and progresses more slowly towards the core, and the evolution of $I_p$ and $l_i(3)$ is accordingly slower. 

\begin{figure}[ht]
	\centering
	\includegraphics[width=100mm]{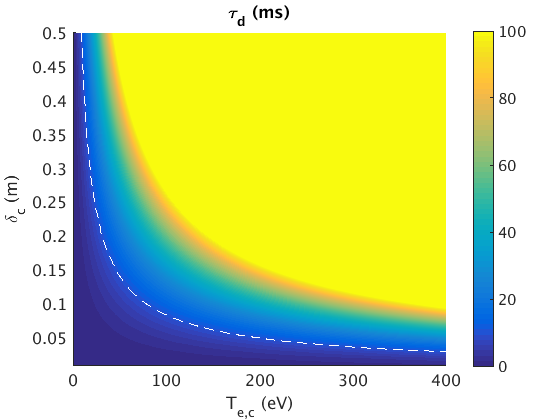}
	\caption{Current diffusion time in the cooled region as a function of its electron temperature $T_{e,c}$ and radial thickness $\delta_c$. The 10 ms contour (corresponding to the shards travel time to the plasma center) is highlighted by the dashed white line.}
	\label{Fig_tau_diff}
\end{figure}

The behaviour displayed by these three simulations may be understood in light of the following considerations. By depositing material, shards cool down the plasma gradually from the edge to the core. Considering that, at any given time, the cooled region is bounded by still hot plasma on one side and by a highly conducting wall on the other side, the current in the cooled region should decay on a current diffusion timescale $\tau_d = \mu_0 \delta_c^2 / \eta_c$, where $\delta_c$ and $\eta_c$ are respectively the radial thickness and the resistivity (assumed homogeneous in this simple, order of magnitude estimate) of the cooled region. Fig. \ref{Fig_tau_diff} shows $\tau_d$ as a function of the electron temperature (assumed homogeneous) in the cooled region $T_{e,c}$ and of $\delta_c$, using the Spitzer resistivity. In the above simulations, $\delta_c$ is obviously increasing in time, but to fix ideas, let us consider that $\delta_c = 0.4$ m, which corresponds to 20 \% of the minor radius. The condition $\tau_d < \tau_{travel} = 10$ ms translates to $T_{e,c} < 12$ eV. This condition is clearly not met in the first simulation (realistic resistivity, no background impurities), in which $T_{e,c}$ remains in the 50 eV range, while it is met in the third simulation, where impurity radiation brings $T_{e,c}$ below 10 eV. This explains why the current profile is not modified in the former case while it is in the latter. In the second simulation, the artifical increase of $\eta$ by a factor 50 corresponds to a proportional decrease of $\tau_d$, such that now $T_{e,c} \simeq 50$ eV is sufficient for a substantial current decay in the cooled region. We have included this simulation because it is common practice in 3D non-linear MHD simulations to use an artifically increased resistivity for numerical reasons. As demonstrated here, this choice may have important consequences, especially in cases where the cold front is `not very cold'. A safer choice is probably to use the Spitzer resistivity up to a certain $T_e$, and then, if necessary, cut-off the $T_e$ dependency of resistivity, i.e. use $\eta = \eta_{Sp}(\min(T_e,T_{e,cut-off}))$. In the present work, we however do not use a cut-off.

An important question is what background impurity density $n_{imp}$ is needed to affect the dynamics. A rough estimate can be drawn by considering that the radiated power per unit volume is $P_{rad} = L_{rad} n_e n_{imp}$ where $L_{rad}$ is the radiative cooling rate. Since the electron thermal energy per unit volume is $e n_e T_e$, the cooling time is $\tau_{cool} = e T_e / (L_{rad} n_{imp}) $, neglecting transport and Ohmic heating and assuming $L_{rad}$ to be a constant. Taking $L_{rad} \simeq 10^{-31}$ W.m$^3$ (which is roughly the peak cooling rate for Argon at coronal equilibrium, see Fig. \ref{Fig_Lrad_imp}) gives $\tau_{cool} \simeq 10^{12} \times T_e/n_{imp}$. In the above simulations, dilution cooling takes $T_e$ down to $\sim 100$ eV. The time it takes for radiation to further reduce $T_e$ to 10 eV or below is thus $\tau_{cool} \simeq 10^{14}/n_{imp}$. The condition $\tau_{cool} > \tau_{travel} = 10$ ms then translates to $n_{imp} < 10^{16}$ m$^{-3}$. JOREK axisymmetric simulations indeed show that $n_{imp} \sim 10^{16}$ m$^{-3}$ is the level above which impurities start to make a difference.

\section{3D simulations}
\label{sec:3D_sims}

We now turn to 3D simulations. We consider only cases with a realistic resistivity. From the above results, it may be expected that cases without background impurities will develop little MHD activity, while cases with $n_{imp}$ above a certain value will develop tearing mode activity due to the effect of shards on the current density distribution. However, the above axisymmetric simulations did not take into account the toroidal localization of the ablation clouds, $\Delta\phi_{cloud}$. We will first study this aspect by scanning $\Delta\phi_{cloud}$ at fixed $n_{imp}$. We will then come back to the role of impurities by scanning $n_{imp}$ at fixed $\Delta\phi_{cloud}$.

\subsection{Effect of the toroidal localization of the ablation clouds}
\label{sec:effect_of_dphicloud}

\begin{figure}[ht]
	\centering
	\includegraphics[width=100mm]{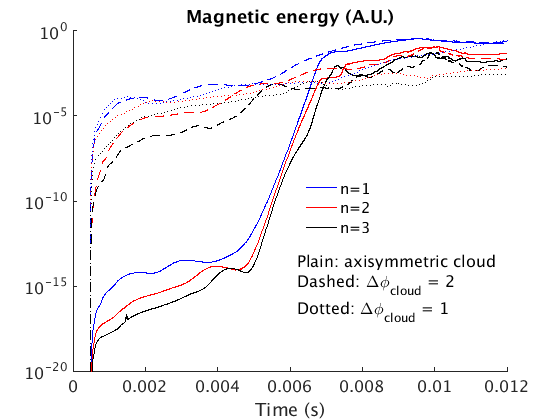}
	\caption{Evolution of the magnetic energy of the $n=1$, 2 and 3 modes for simulations with $\Delta\phi_{cloud} = 1000$, 2 and 1 rad (all with realistic resistivity and $n_{imp} = 10^{17}$ m$^{-3}$)}
	\label{DPHI1000_vs_2_vs_1_modes_energy}
\end{figure}

Fig. \ref{DPHI1000_vs_2_vs_1_modes_energy} compares the evolution of the magnetic energy of the  $n=1$ (blue), 2 (red) and 3 (black) modes for three values of $\Delta\phi_{cloud}$ (in radians): 1000 i.e. virtually axisymmetric clouds (plain lines), 2 (dashed lines) and 1 (dotted lines), all other parameters being the same. These simulations include background impurity radiation with $n_{imp} = 10^{17}$ m$^{-3}$. Two phases can be distinguished. Before $\sim 5$ ms, modes are many orders of magnitude weaker for the $\Delta\phi_{cloud} = 1000$ case than for the other ones. After 5 ms, modes grow strongly for the $\Delta\phi_{cloud} = 1000$ case and the mode amplitudes become comparable to the $\Delta\phi_{cloud} = 2$ case around 7 ms and then saturate. The $n=1$ mode is dominant. The $\Delta\phi_{cloud} = 1$ case, although it has comparable $n=1$ and larger $n=2$ and 3 mode amplitudes to the $\Delta\phi_{cloud} = 2$ case in the early phase (and thus a broader $n$ spectrum, consistently with the smaller $\Delta\phi_{cloud}$), has a much smaller $n=1$ amplitude after 5 ms (until 10 ms). Thus, it appears that the drive of low $n$ modes decreases (or at least is delayed) as the ablation cloud becomes more strongly toroidally localized. 

\begin{figure}[ht]
	\centering
	\includegraphics[trim=0 40 0 40, clip, width=100mm]{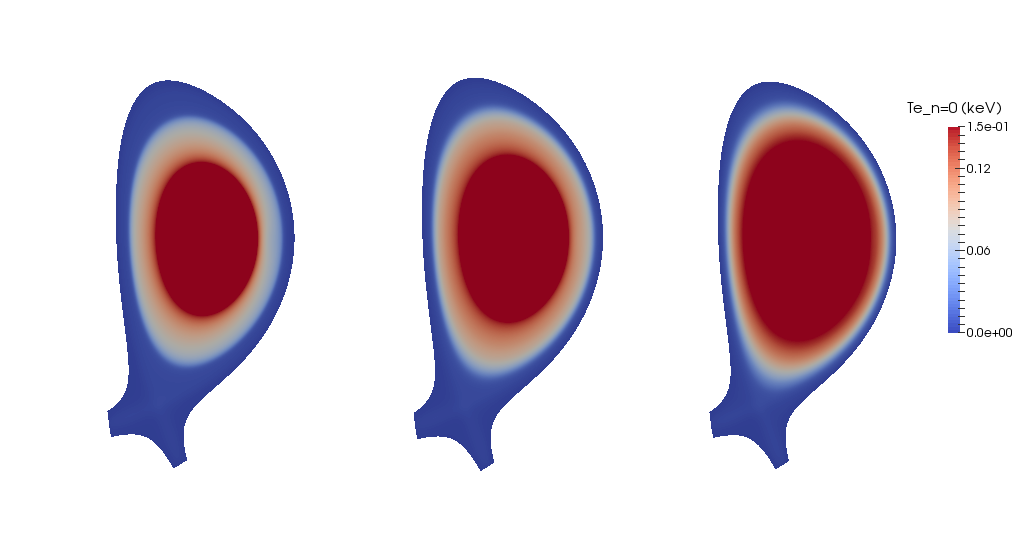}
	\includegraphics[trim=0 40 0 40, clip, width=100mm]{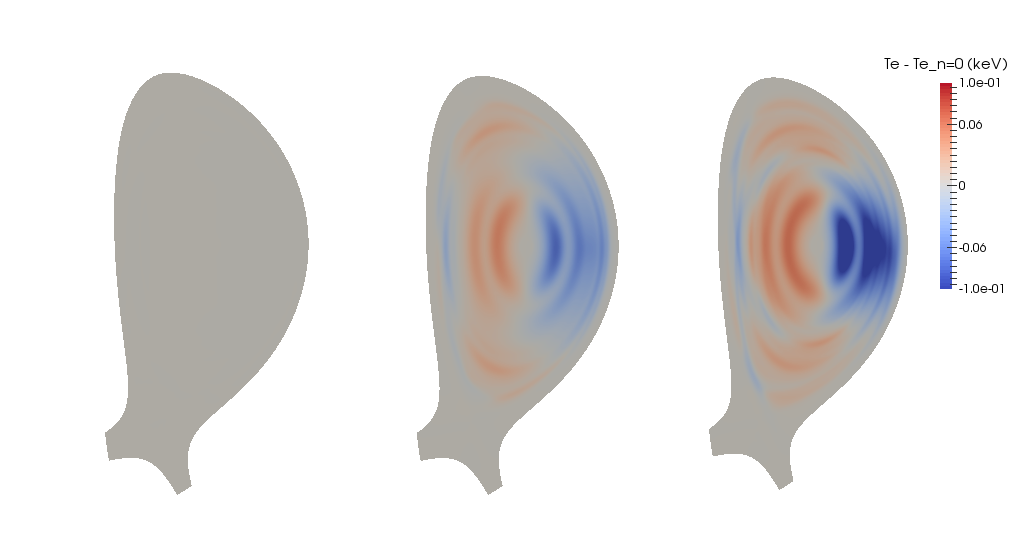}
	\caption{Axisymmetric component (top) and deviation from axisymmetry (bottom) of the electron temperature at t=4 ms for $\Delta\phi_{cloud} = 1000$ (left),  $\Delta\phi_{cloud} = 2$ (middle) and $\Delta\phi_{cloud} = 1$ rad (right) (all with realistic resistivity and $n_{imp} = 10^{17}$ m$^{-3}$). In the top row, the colourscale is saturated at 150 eV so as to highlight differences in the edge, but the core temperature is much higher, around 5 keV.}
	\label{DPHI1000_vs_2_vs_1_cuts_4ms}
\end{figure}

Fig. \ref{DPHI1000_vs_2_vs_1_cuts_4ms} helps interpret these results. It presents poloidal cross-sections of the axisymmetric part of the electron temperature, $T_{e,n=0}$ (top row), and of its deviation from axisymmetry, $T_e - T_{e,n=0}$ (bottom row), for the three simulations at $t=4$ ms. It can be seen, not surprisingly, that the effect on $T_{e,n=0}$ is larger for larger $\Delta\phi_{cloud}$ while the effect on $T_e - T_{e,n=0}$ is larger for smaller $\Delta\phi_{cloud}$. As has been discussed in previous publications \cite{Nardon_PPCF_2017}\cite{Hu_NF_2018}, there are two main ways through which tearing modes can be driven by a massive material injection: the first one is \textit{via} a modification of the current density profile leading to a positive $\Delta'$, and the second one is \textit{via} helical cooling of flux tubes on rational surfaces. Only the first mechanism is present in the $\Delta\phi_{cloud} = 1000$ case. In the early phase, mode excitation is dominated by the second mechanism, explaining the much smaller mode amplitude in the $\Delta\phi_{cloud} = 1000$ case than in the other ones. From about 5 ms, a fast and large mode growth takes place for the $\Delta\phi_{cloud} = 1000$ case as the first mechanism is strongly activated by the skin current passing across the $q=2$ surface (note that the early evolution of profiles for this simulation is very similar to that of the third axisymmetric simulation, i.e. to the black profiles in Fig. \ref{Fig_axisym_profiles}). Fig. \ref{DPHI1000_vs_2_vs_1_cuts_4ms} shows that decreasing $\Delta\phi_{cloud}$ reduces the first mechanism and increases the second one. These two changes roughly compensate each other for the $\Delta\phi_{cloud} = 2$ case, which has a comparable maximal $n=1$ amplitude to the $\Delta\phi_{cloud} = 1000$ case. On the other hand, for the $\Delta\phi_{cloud} = 1$ case, the much smaller $n=1$ amplitude shows that the reduction of the $\Delta'$ effect is not compensated by the increase of the helical cooling effect. The detailed reasons why a smaller $\Delta\phi_{cloud}$ leads to a reduction of the axisymmetric temperature (and current) profile modification would deserve a specific study, but are probably related to the fact that localized ablation clouds promptly generate magnetic islands at rational $q$ surfaces, whose O-points coincide with their position. The cooling then tends to remain localized inside the islands and thus non-axisymmetric, which limits axisymmetric profile modifications. Another contributing factor is that, due to a stronger local reduction of the temperature, the ablation rate decreases with $\Delta\phi_{cloud}$. At $t =$ 4 ms, the number of ions in the plasma is $3.5 \times 10^{23}$, $2.9 \times 10^{23}$ and $2.2 \times 10^{23}$ for $\Delta\phi_{cloud} = 1000$, 2 and 1 respectively, showing that this effect is substantial.

\subsection{Effect of background impurities}

\begin{figure}[ht]
	\centering
	\includegraphics[width=60mm]{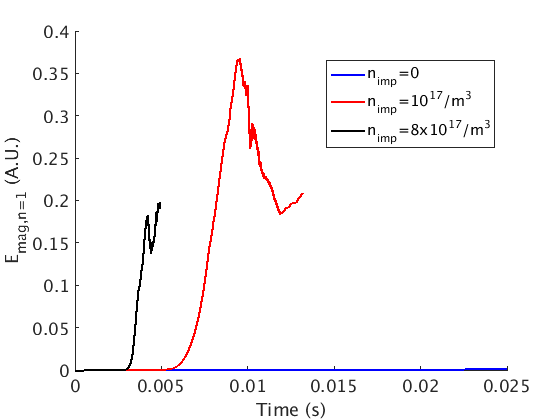}
	\includegraphics[width=60mm]{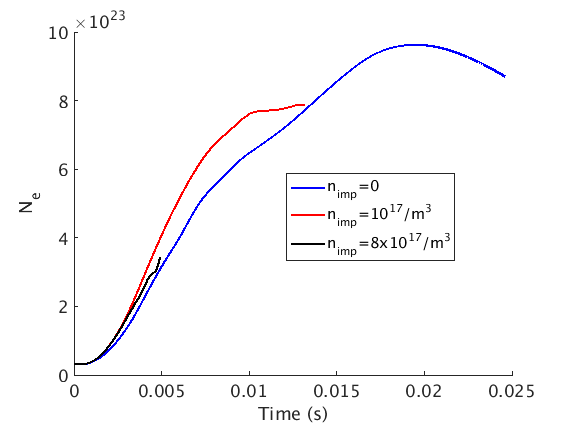}
	\includegraphics[width=60mm]{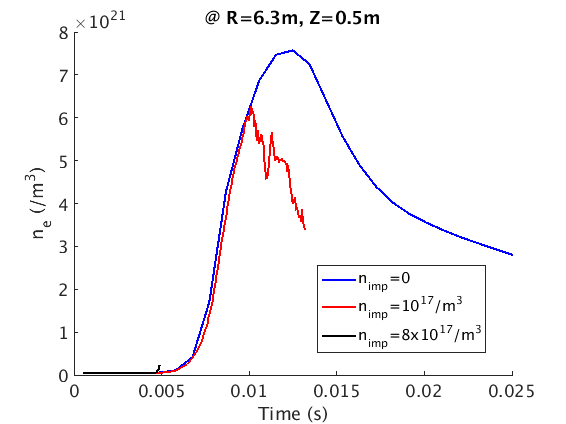}	
	\includegraphics[width=60mm]{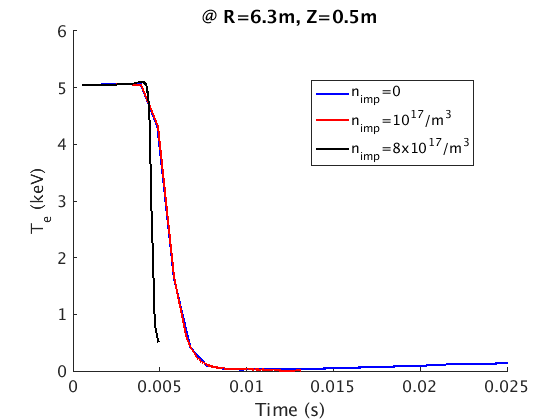}	
	\caption{Time evolution of the magnetic energy of the $n=1$ mode (top left), the total number of free electrons in the plasma (top right), the central electron temperature (bottom left) and the central electron density (bottom right) for simulations with $n_{imp}=0$ (blue), $10^{17}$ m$^{-3}$ (red) and $8 \times 10^{17}$ m$^{-3}$ (black)}
	\label{fig:nimp_scan}
\end{figure}

\begin{figure}[ht]
	\centering
	\includegraphics[trim=130 0 140 0, clip, width=30mm]{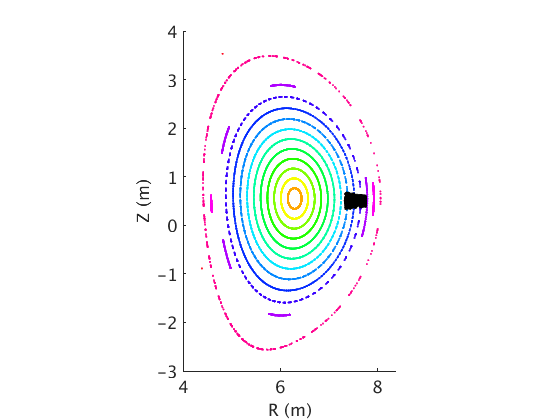}
	\includegraphics[trim=130 0 140 0, clip, width=30mm]{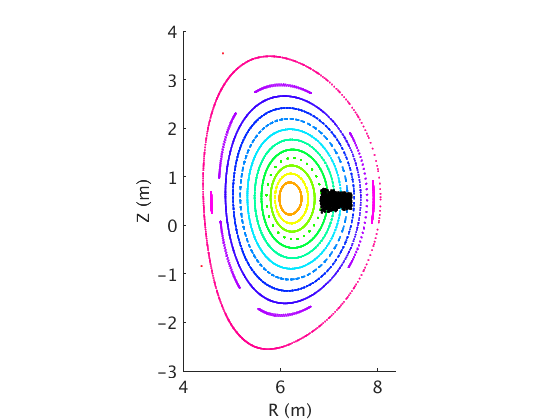}	
	\includegraphics[trim=130 0 140 0, clip, width=30mm]{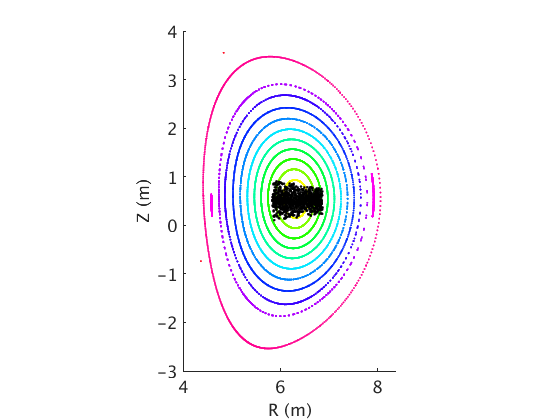}		
	\includegraphics[trim=130 0 140 0, clip, width=30mm]{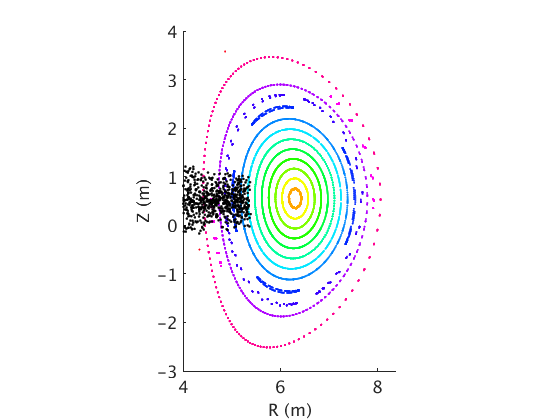}		
	\caption{Poincar\'e cross-sections at t=4, 6, 10 and 20 ms (from left to right) for the simulation with $n_{imp}=0$. The shards position is shown in black.}
	\label{fig:Poinc_nimp0}
\end{figure}

We now come back to studying the effect of background impurities, but this time with 3D simulations. We choose $\Delta\phi_{cloud} = 2$ as a `pessimistic' value in the sense that, as shown above, low $n$ modes (which are responsible for the TQ triggering \cite{Nardon_PPCF_2017}) are more strongly driven for this value than for a smaller, more realistic $\Delta\phi_{cloud}$. Fig. \ref{fig:nimp_scan} compares the magnetic energy of the $n=1$ mode (top left), the total number of free electrons in the plasma, $N_e$ (top right), the central $T_e$ (bottom left) and the central $n_e$ (bottom right) for simulations with $n_{imp}=0$ (blue), $10^{17}$ m$^{-3}$ (red) and $8 \times 10^{17}$ m$^{-3}$ (black), all other parameters being equal. It can be seen that the $n=1$ mode is substantially excited only in the presence of impurities. Fig. \ref{fig:Poinc_nimp0} shows a series of Poincar\'e cross-sections at different times for the $n_{imp}=0$ simulation, with the position of shards overlayed in black, revealing that shards travel across the plasma without perturbing flux surfaces much. By $t=20$ ms, when shards have almost left the plasma, about $10^{24}$ free electrons have been added to the plasma (see the top right plot of Fig. \ref{fig:nimp_scan}). This represents half of the initial pellet content. The other half splits roughly equally between leftover shards and transport losses.

\begin{figure}[ht]
	\centering
	\includegraphics[trim=130 0 140 0, clip, width=30mm]{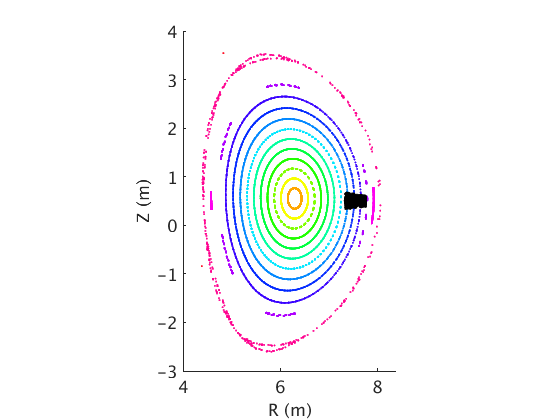}
	\includegraphics[trim=130 0 140 0, clip, width=30mm]{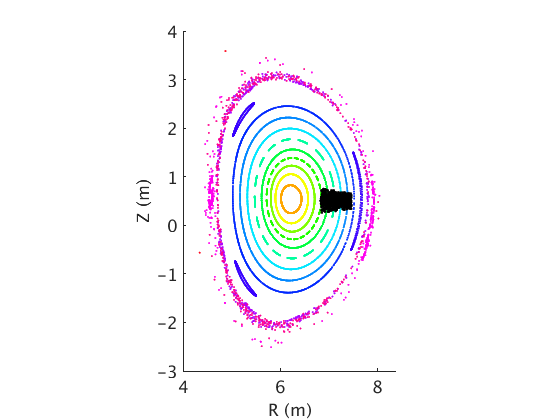}	
	\includegraphics[trim=130 0 140 0, clip, width=30mm]{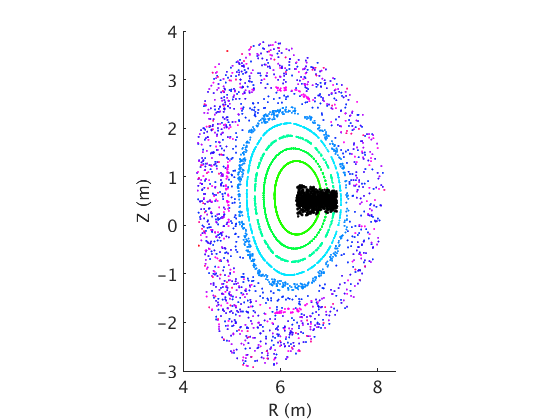}		
	\includegraphics[trim=130 0 140 0, clip, width=30mm]{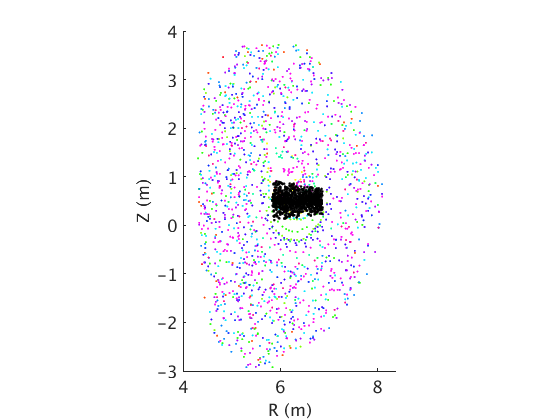}		
	\caption{Poincar\'e cross-sections at t=4, 6, 8 and 10 ms (from left to right) for the simulation with $n_{imp}=10^{17}$ m$^{-3}$. The shards position is shown in black.}
	\label{fig:Poinc_nimp1e17}
\end{figure}

For $n_{imp} = 10^{17}$ m$^{-3}$, the $n=1$ mode is excited from about 6 ms. The series of Poincar\'e cross-sections shown in Fig. \ref{fig:Poinc_nimp1e17} reveal the associated destruction of flux surfaces and growth of a stochastic region from the edge to the core. It can be seen that shards travel faster than the inner boundary of the stochastic region up to about the time when they reach the core, at 10 ms, at which point the magnetic field is fully stochastized. The evolution of the core $n_e$ and $T_e$ is similar to that observed in the $n_{imp} = 0$ case until 10 ms, consistently with the persistence of flux surfaces in the core region until this time.

\begin{figure}[ht]
	\centering
	\includegraphics[trim=130 0 140 0, clip, width=30mm]{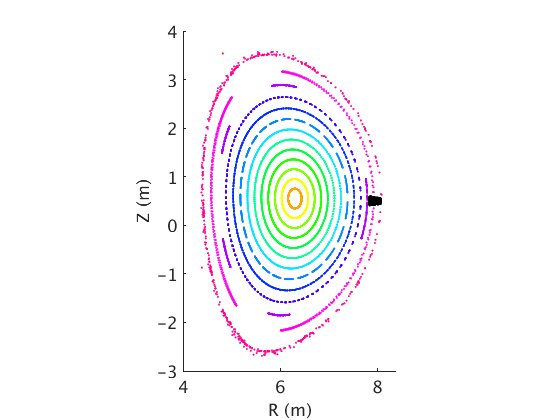}	
	\includegraphics[trim=130 0 140 0, clip, width=30mm]{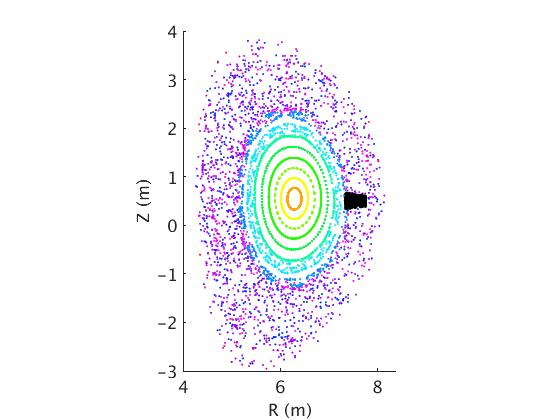}
	\includegraphics[trim=130 0 140 0, clip, width=30mm]{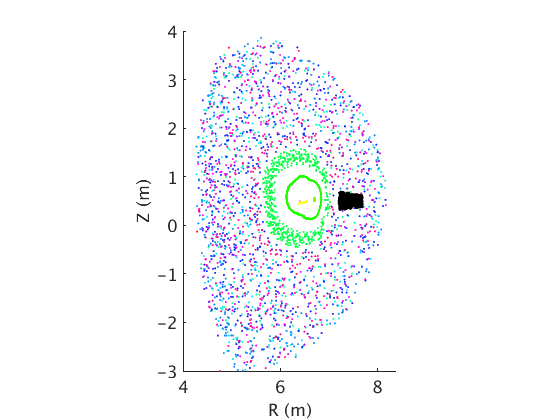}	
	\includegraphics[trim=130 0 140 0, clip, width=30mm]{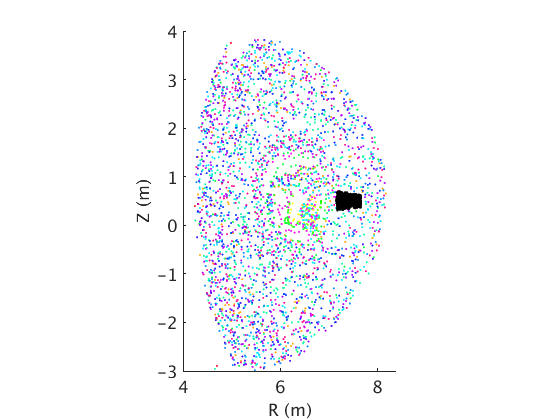}		
	\caption{Poincar\'e cross-sections at t=2, 4, 4.5 and 4.7 ms (from left to right) for the simulation with $n_{imp}=8 \times 10^{17}$ m$^{-3}$. The shards position is shown in black.}
	\label{fig:Poinc_nimp8e17}
\end{figure}

For $n_{imp} = 8 \times 10^{17}$ m$^{-3}$, the $n=1$ mode is excited earlier, from about 3 ms. This is because the larger $n_{imp}$ results in a shorter duration for the radiative collapse to take place in the cold front. The series of Poincar\'e cross-sections shown in Fig. \ref{fig:Poinc_nimp8e17} now show that the stochastic region expands faster than the shards velocity. As a result, the magnetic field becomes fully stochastic while the shards are still located around mid-radius. This stochastization causes a fast drop of the core $T_e$ before $n_e$ has substantially increased in this region (see the bottom plots of Fig. \ref{fig:nimp_scan}): this is an illustration of a `premature' TQ, which may be problematic regarding RE generation. The simulation is not pushed further since the purpose of this work is not to study TQ dynamics, but only to investigate whether or not a TQ is triggered.


\section{Discussion and conclusion}
\label{sec:discussion}

The above simulations suggest that for the considered ITER 15 MA L-mode scenario, pure D$_2$ SPI may be able to substantially fuel the plasma all the way to the core without provoking a major MHD-driven TQ, provided the background impurity density is low enough. It may be speculated that a similar conclusion would have been drawn for an H-mode case since, according to the considerations made in Section \ref{sec:axisym_sims}, hotter plasmas should be harder to destabilize than colder ones (for a given dilution factor). On the other hand, H-mode plasmas could be much more sensitive to edge instabilities, which might change the picture. This question should be addressed by future simulations. Another important point is that, although our simulations strongly exaggerate the size of ablation clouds, results presented in Section \ref{sec:effect_of_dphicloud} suggest that more localized ablation clouds would reinforce our conclusion.

\begin{figure}[ht]
	\centering
	\includegraphics[width=100mm]{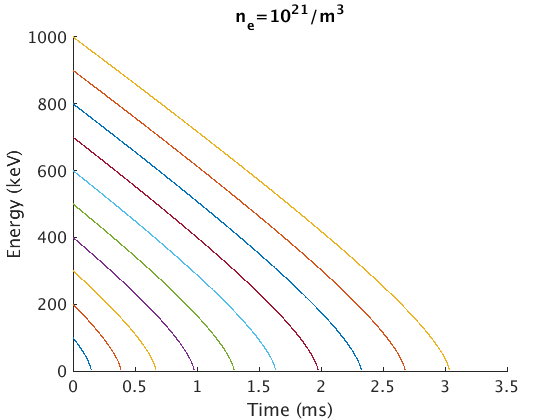}
	\caption{Electron energy versus time, for different initial energies, for test electrons undergoing only a friction force from bulk electrons}
	\label{Fig_electron_slowing_down}
\end{figure}

Overall, these results motivate the consideration and further exploration of a disruption mitigation strategy based on a pure D$_2$ SPI followed by impurity-containing SPI. As mentioned in the introduction, this is probably the most efficient way to fuel the plasma and thus to implement the present RE avoidance strategy in ITER \cite{Martin-Solis_NF_2017}. An important additional observation is that the relatively slow dilution cooling could be extremely efficient at suppressing risks of hot tail RE generation during the TQ. Fig. \ref{Fig_electron_slowing_down} shows how the energy of an electron evolves under pure drag from the bulk electrons. It has been obtained by solving the equation $dp/dt = -e E_{CH} c^2/v^2$ where $p$ is the electron momentum and $v$ its velocity, $E_{CH}$ is the Connor-Hastie field and $c$ is the speed of light (see e.g. Appendix A in  \cite{Boozer_NF_2017}). A bulk electron density of $10^{21}$ m$^{-3}$ has been assumed. It can be seen that even at an initial energy of 1 MeV, electrons are braked within 3 ms. Thus, if $n_e$ could be increased to above $10^{21}$ m$^{-3}$ and maintained at this level for at least 3 ms before the TQ takes place, the electron population would remain Maxwellian, and the moderate post-dilution $T_e$ would probably remove the risk of hot tail RE generation. The above JOREK simulation results suggest that this is feasible, provided the background impurity density is low enough. 

The most serious threat to the proposed strategy indeed appears to be the presence of background impurities. In this study, we have considered an Argon-like impurity and have found that for $n_{imp}$ larger than $\sim 10^{17}$ m$^{-3}$, there is a risk to trigger a TQ before shards have reached the plasma center. Future work should aim for a more realistic treatment of background impurities. We note that the NIMROD study on uniform D$_2$ delivery had considered Beryllium as the background impurity with a density of $5 \times 10^{18}$ m$^{-3}$, and found no $n=1$ or $n=2$ mode excitation over the simulated duration of 12 ms, in spite of the large dilution factor of 150 \cite{Izzo_2009}. The effect of pre-existing magnetic islands could also represent a threat to the proposed strategy and will be the object of upcoming work. 

Yet another point to consider is the fact that a substantial fraction of gas may be produced during the shattering process. As shown in Fig. 6 of \cite{Gebhart_2020}, above a certain pellet velocity, the mass fraction contained in solid fragments drops sharply from $> 70$ \% to $< 10$ \%. A large gas fraction is clearly not desirable since the gas will not fuel the plasma directly in the core but will instead fuel at the edge \cite{Nardon_NF_2017}, with the risk of cooling down the edge too much and triggering MHD modes \cite{Nardon_PPCF_2017}. This sets a limit on the pellet velocity, which depends on the angle of the shattering tube. Future work should be devoted to optimizing the injector design, taking into account the gas cloud accompanying or preceding the solid shards and a more realistic description of the shard plume characteristics based on existing knowledge \cite{Gebhart_2020}\cite{Lehnen_IAEA_TM_2020}.

Future efforts should also be devoted to the validation of pure D$_2$ SPI simulations on present experiments, e.g. on DIII-D, JET, KSTAR or ASDEX Upgrade, with however the expected difficulty that the behaviour may be strongly influenced by the presence of background impurities whose properties are difficult to assess.

Finally, it would be useful to test the proposed 2-step SPI scheme experimentally in present machines equipped with multiple (shattered) pellet injection systems, or even using SPI for the first (D$_2$) injection followed by MGI for the second (impurity) injection.

\section{Acknowledgements}

Parts of this work have been carried out within the framework of the EUROfusion Consortium and have received funding from the Euratom Research and Training Programme 2014-2018 and 2019-2020 under Grant Agreement No. 633053. Part of this work was carried out using the Marconi-Fusion supercomputer. The views and opinions expressed herein do not necessarily reflect those of the European Commission. Part of this work is supported by the National Natural Science Foundation of China under Grant No. 11905004.



\end{document}